\documentclass[10pt, twocolumn, pre, aps, superscriptaddress, showpacs]{revtex4-1}
\usepackage{amsmath, graphicx, subfigure, tikz}
\begin{document}

\title{Spin-s Spin-Glass Phases in the d=3 Ising Model}
    \author{E. Can Artun}
    \affiliation{Faculty of Engineering and Natural Sciences, Kadir Has University, Cibali, Istanbul 34083, Turkey}
    \author{A. Nihat Berker}
    \affiliation{Faculty of Engineering and Natural Sciences, Kadir Has University, Cibali, Istanbul 34083, Turkey}
    \affiliation{Department of Physics, Massachusetts Institute of Technology, Cambridge, Massachusetts 02139, USA}

\begin{abstract}

All higher-spin $(s\geq 1/2)$ Ising spin glasses are studied by renormalization-group theory in spatial dimension $d=3$, exactly on a $d=3$ hierarchical model and, simultaneously, by the Migdal-Kadanoff approximation on the cubic lattice.  The s-sequence of global phase diagrams, the chaos Lyapunov exponent, and the spin-glass runaway exponent are calculated. It is found that, in $d=3$, a finite-temperature spin-glass phase occurs for all spin values, including the continuum limit of $s\rightarrow \infty$.  The phase diagrams, with increasing spin $s$, saturate to a limit value.  The spin-glass phase, for all $s$, exhibits chaotic behavior under rescalings, with the calculated Lyapunov exponent of $\lambda = 1.93$ and runaway exponent of $y_R=0.24$, showing simultaneous strong-chaos and strong-coupling behaviors.  The ferromagnetic-spinglass and spinglass-antiferromagnetic phase transitions occurring, along their whole length, respectively at $p_t = 0.37$ and 0.63 are unaffected by $s$, confirming the percolative nature of this phase transition.

\end{abstract}
\maketitle

\section{Introduction: Spin-s Ising Spin-Glass Systems}

Since the blossoming days of modern phase transitions and critical phenomena, theoretical insight has been obtained from model spin systems studied systematically as a function of the number of local states, namely for all $s$.\cite{spinS1,spinS2,spinS3,spinS4,spinS5,spinS6,spinS7,spinS8,spinS9,spinS10,spinS11,spinS12,spinS13,spinS14,spinS15}. Indeed some of the authors in these References are the founders of the field.  However, the same systematic studies have not been done on the complex ordering systems of current interest, such as in the presence of frozen disorder, presumably due to calculational difficulties, now surmountable by grobal renormalization group, as shown below.

Frozen disorder of the interactions introduces many qualitatively and quantitatively new effects to statistical mechanical systems, such as the immediate (i.e., with infinitesimal disorder) conversion of first-order phase transitions into second-order phase transitions \cite{Aizenman,AizenmanE,HuiBerker,erratum} or the creation of an entirely new phase such as the spin-glass phase \cite{EdwardsAnderson}.  The latter occurs under frozen (quenched) competing interactions causing local minimum-energy degeneracies dubbed frustration \cite{Toulouse}.  The signature of the spin-glass phase is the appearance of a chaotic sequence of interactions \cite{McKayChaos,McKayChaos2,BerkerMcKay,Hartford,ZZhu,Katzgraber3,Fernandez,Fernandez2,Eldan,Wang2,Parisi3} under the successive scale changes of a renormalization-group transformation. This translates to a chaotic spin-spin correlation function, as function of distance, at a given scale.\cite{Aral} This chaotic behavior signifies that a small change at the microscopic level in the coupling constant $(J)$ or temperature $(1/J)$, or randomness $(p)$ in the system causes major changes in the correlation function between two distant spins.\cite{Aral}

\begin{figure}[ht!]
\centering
\includegraphics[scale=0.45]{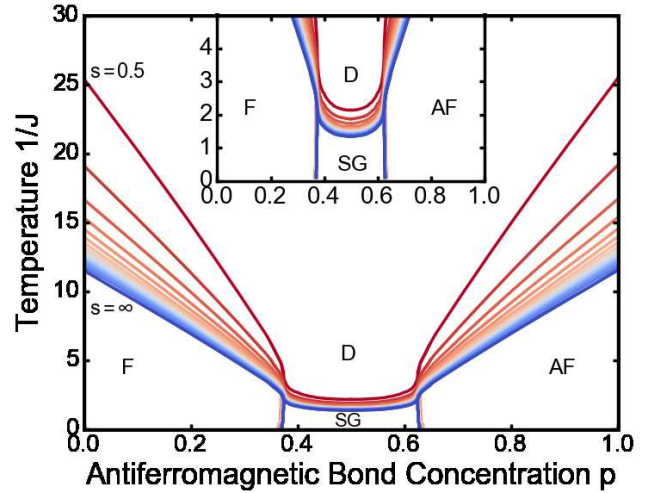}
\caption{Calculated phase diagrams of the spin-$s$ Ising spin glasses in $d=3$. Fom top to bottom, $s=1/2,1,3/2,2,5/2,3,...$ to $s\rightarrow \infty$. There is an accumulation, from above, of the phase diagrams at the lowermost, but still at finite-temperature, phase diagram of the continuum limit $s\rightarrow \infty$. The phase diagrams shown here are the results of our numerical calculations and are accurate to the thickness of the lines in this figure.}
\end{figure}

The spin-glass phase and its rescaling chaos appears with the introduction, by rewiring, of infinitesimal frustration to the Mattis phase \cite{Mattis} obtained by random local spin redefinitions (gauge transformations) in the usual ferromagnetic or antiferromagnetic phase.\cite{Ilker2}  On the other hand, strong chaos, signalled by a large Lyapunov exponent, of the spin-glass phase in fully frustated systems continues \cite{Demirtas} until the lower-critical dimension $d_c\simeq 2.5$ of the spin-glass phase \cite{Parisi,Boettcher,Parisi2,Amoruso,Bouchaud,Demirtas,Atalay}. Thus both gradual \cite{Ilker2} or abrupt \cite{Demirtas} onsets of chaos are seen.

Most spin-glass studies have been on the classical spin $s = 1/2$ Ising model, where locally $s_i = \pm 1$.\cite{Binder}  Spin-glass studies have also been done on $q$-state clock models and their continuum limit the XY model \cite{Ilker1,Ilker3}, chiral (helical \cite{Surface3}) Potts and clock models, in fact leading to a chiral spin-glass Potts \cite{Caglar1} and clock \cite{Caglar2, Caglar3} phases, and quantum Heisenberg models \cite{Kaplan}.  The position-space renormalization-group method appears to be a method suited for such studies, where the rescaling behavior of the distribution of the quenched random interactions is followed and analyzed \cite{AndelmanBerker}.  This is best effected (Fig. 2) by use of the Migdal-Kadanoff approximation \cite{Migdal,Kadanoff} or, equivalently, the exact recursion of a hierarchical lattice \cite{BerkerOstlund,ArtunBerker,Kaufman1,Kaufman2}.  In the current work, we quantitatively and globally study, in spatial dimension $d=3$, the Ising spin glass for all spins $s= 1/2,1,3/2,2,5/2,...$ to the limiting value $s\rightarrow \infty$, obtaining the global $s$-sequence phase diagram (Fig. 1) and chaotic behaviors.

The spin-$s$ Ising model is defined by the Hamiltonian
\begin{equation}
-\beta \mathcal{H}=\sum_{\langle ij \rangle} J_{ij} (s_i/s) (s_j/s) \,,
\end{equation}
where $\beta=1/kT$, at each site $i$ of the lattice the spin $s_i =
\pm1/2,\pm1,\pm3/2,...,\pm s$, and $\langle ij \rangle$ denotes summation over all
nearest-neighbor site pairs.  The division by $s$ is done to conserve the energy scale across the different spin-$s$ models and thereby make meaningful temperature comparisons between them.  Note that for $s=1/2$, this formalism yields the much studied $s_i/s = \pm1$ case. The bond $J_{ij}$ is ferromagnetic $+J>0$ or antiferromagnetic $-J$ with respective probabilities $1-p$ and $p$. Under renormalization-group tranformation, this "double-delta" distribution of interactions is not conserved. A more complicated distribution of interactions ensues and is kept track of, as explained below.

\begin{figure}[ht!]
\centering
\includegraphics[scale=0.9]{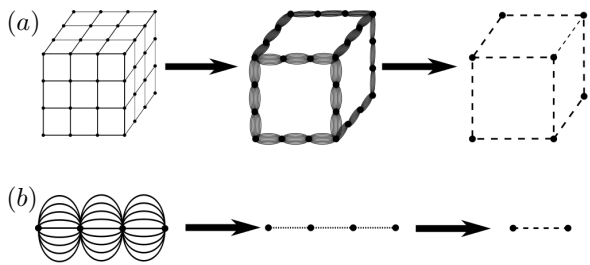}
\caption{(a) Migdal-Kadanoff approximate renormalization-group transformation for the $d=3$ cubic lattice with the length-rescaling factor of $b=3$. In this intuitive approximation, bond moving is followed by decimation. (b) Exact renormalization-group transformation of the $d=3, b=3$ hierarchical lattice for which the Migdal-Kadanoff renormalization-group recursion relations are exact. The construction of a hierarchical lattice proceeds in the opposite direction of its renormalization-group solution. From \cite{BerkerOstlund,Caglar2}.}
\end{figure}

\section{Method: Renormalization-Group Flows of the Quenched Probability Distribution of the Interactions}

Under renormalization group, for $s>1/2$, the Hamiltonian does not conserve its form in Eq.(1).  Thus, for any $s$, the Hamiltonian is most generally expressed as
\begin{equation}
- \beta {\cal H} =  \sum_{\left<ij\right>} E_{ij}(s_i,s_j) \, ,
\end{equation}
where $E_{ij}(s_i,s_j)$ is the interaction between nearest-neighbor spins, starting out as in Eq.(1) as $ J_{ij} (s_i/s) (s_j/s)$ and generalizing to a $(2s+1) \times (2s+1)$ matrix under renormalization group.  With no loss of generality, for each $<ij>$, the same constant is subtracted from all terms $E(s_i,s_j)$, so that the largest energy $E(s_i,s_j)_{max}$ of the spin-spin interaction is zero (and all other $E(s_i,s_j) < 0$). This formulation makes it possible to follow global renormalization-group trajectories, necessary for the calculation of phase boundaries, Lyapunov exponent, and runaway exponent, without running into numerical overflow problems.  As the local renormalization-group transformation, the Migdal-Kadanoff approximate transformation \cite{Migdal,Kadanoff,Biroli} and, equivalently, the exact transformation for the $d=3$ hierarchical lattice \cite{BerkerOstlund,Kaufman1,Kaufman2} is used (Fig. 2). Recent works using exactly soluble hierarchical models are in Refs.
\cite{Myshlyavtsev,Derevyagin,Shrock,Monthus,Sariyer,Ruiz,Rocha-Neto,Ma,Boettcher5}. The length rescaling factor of $b=3$ is used, to preserve under renormalization group the ferromagnetic-antiferromagnetic symmetry of the system.  This local transformation consists in bond moving followed by decimation, with the above-mentioned subtraction after each local bond moving and decimation, giving the local renormalized energies $E'(s_i,s_j) \leq 0$.  In our notation, all renormalized quantities are designated by a prime.

The quenched randomness is included by keeping, as a distribution, 10000 sets of the nearest-neighbor interaction energies ${E(s_i,s_j)}$.  At the beginning of each renormalization-group trajectory, this distribution is formed from the double-delta distribution characterized by interactions $\pm J$ with probabilities $p,(1-p)$.  10000 local renormalization-group transformations determine each subsequent distribution as explained below.

The local renormalization-group transformation is simply expressed in terms of the transfer matrix $T(s_i,s_j)= e^{E(s_i,s_j)}$:  Bond moving consists of multiplying elements at the same position of $b^{d-1}=9$ transfer matrices randomly chosen from the distribution,
\begin{equation}
\widetilde{T}(s_i,s_j) =  \prod_{k=1}^9 T_k(s_i,s_j) \, ,
\end{equation}
so that a distribution of 10000 bond-moved transfer matrices is generated. Decimation consists of matrix multiplication of three randomly chosen bond-moved transfer matrices,
\begin{equation}
\mathbf{T'} = \mathbf{\widetilde{T}_1\cdot\widetilde{T}_2\cdot\widetilde{T}_3} \, ,
\end{equation}
so that a distribution of 10000 renormalized transfer matrices is generated.  Phases are determined by following trajectories to their asymptotic limit:  The asymptotic limit transfer matrices of trajectories starting in the ferromagnetic phase all have 1 in the corner diagonals and 0 at all other positions.  The asymptotic limit transfer matrices of trajectories starting in the antiferromagnetic phase all have 1 in the corner anti-diagonals and 0 at all other positions.  The asymptotic limit transfer matrices of trajectories starting in the spin-glass phase all have 1 in the corner diagonals $(s_i=s,s_j=s)$ and $(s_i=-s,s_j=-s)$ or in the corner anti-diagonals $(s_i=s,s_j=-s)$ and $(s_i=-s,s_j=s)$, and 0 at all other positions.  We use $(s,s-1,s-2,...,-s+2,-s+1,-s)$ for the order of the rows and columns in the matrix $E(s_i,s_j)$. The asymptotic limit transfer matrices of trajectories starting in the disordered phase all have 1 at all other positions.  Phase diagrams are obtained by numerically determining the boundaries, in the unrenormalized system, of these asymptotic flows.  Thus, to numerically obtain the phase diagram, we determine the asymptotic limit, under repeated renormalization-group transformations, of the distribution of transfer matrices.  These asymptotic limits for the ferromagnetic, antiferromagnetic, spin-glass, and disordered phases are distinct, as described here.  The boundaries between different asymptotic limits, as a function of starting (unrenormalized) $J$ and $p$, yield the phase diagram numerically, as seen in Fig. 1 for different values of the spin $s$.

\section{Results: Global s-Sequence Phase Diagram and Saturation}

\begin{figure}[ht!]
\centering
\includegraphics[scale=0.45]{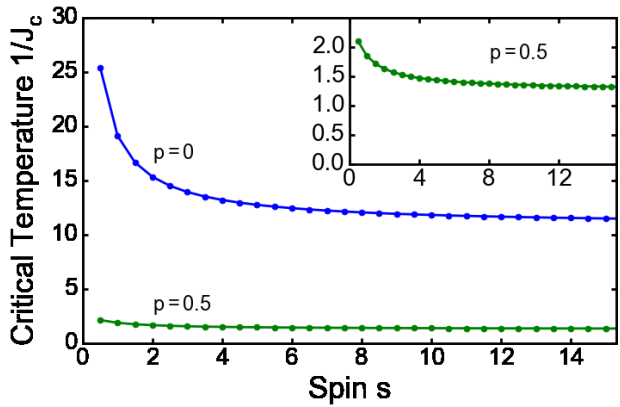}
\caption{The calculated ferromagnetic (at $p=0$) and spin-glass (at $p=0.5$) phase transition temperatures as a function of spin value $s$.  Note that with increasing $s$ both transition temperatures saturate around $s\simeq 8$.  A similar behavior was found in $q$-state clock models.\cite{ArtunBerker} The critical temperatures in this figure are the results of our numerical calculations and are accurate to the size of the data points in this figure.}
\end{figure}

\begin{figure}[ht!]
\centering
\includegraphics[scale=0.4]{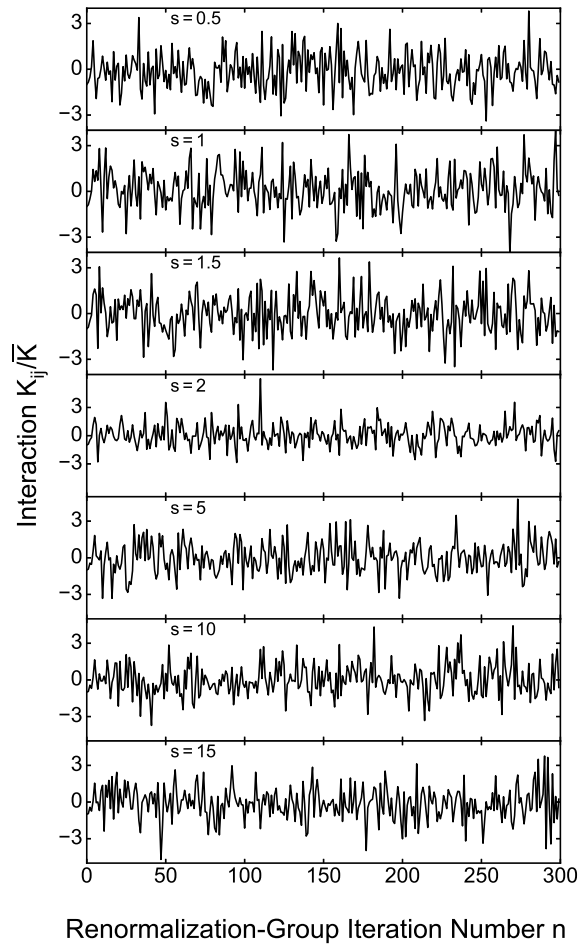}
\caption{The chaotic renormalization-group trajectory of the interaction $K_{ij}$ at a given location $<ij>$, for various spin $s$
values, at spatial dimension $d = 3$. Note the strong chaotic behavior for all $s$, as also reflected by the calculated Lyapunov exponent $\lambda =1.93$ for all $s$.  The calculated runaway exponent is $y_R=0.24$ for all $s$, showing simultaneous strong-chaos and strong-coupling behaviors.}
\end{figure}

In this and the next Sections, we present the numerical results on spin-$s$ Ising spin-glass systems, calculated exactly on a $d=3$ hierarchical model and, simultaneously, by the Migdal-Kadanoff approximation on the cubic lattice.

The calculated phase diagrams of the spin-$s$ Ising spin glasses in $d=3$ are shown in Fig. 1. Fom top to bottom, the phase diagrams are for spin-$s=1/2,1,3/2,2,5/2,3,...$ to $s\rightarrow \infty$. There is an accumulation, from above, of the phase diagrams at the lowermost, but still at finite-temperature, phase diagram of the continuum limit $s\rightarrow \infty$.  The phase diagrams shown in Fig.1 are the results of our numerical calculations and are accurate to the thickness of the lines in this figure.

The calculated ferromagnetic (at $p=0$) and spin-glass (at $p=0.5$) phase transition temperatures as a function of spin value $s$ are given in Fig. 3.  With increasing $s$ both transition temperatures saturate around $s\simeq 8$.  A similar behavior was found in $q$-state clock models saturating at the continuum XY model transition temperature.\cite{ArtunBerker} The critical temperatures in Fig. 3 are the results of our numerical calculations and are accurate to the size of the data points in this figure.

\section{Results: Chaos for All Spins s, Lyapunov Exponent and Runaway Exponent}

For all spin-$s$, the renormalization-group trajectories starting within the spin-glass phase are asymptotically chaotic, as seen in Fig. 4, where the consecutively renormalized (combining with neighboring interactions) values at a given location, meaning $<ij>$ then the renormalized $<i'j'>$ that includes $<ij>$, etc., are followed.  For the interaction $K_{ij}$, we have used the difference between the largest value (which is 0 by construction) and the lowest value in $E(s_i,s_j)$. $\overline{K}$ is the average of this interaction over the entire distribution at the given renormalization-group step. Since dimensionless interaction $K$ is proportional to inverse temperature as seen from our Eq.(1), chaotic K means chaotic temperature.

The chaotic behavior is strong, as measured by the Lyapunov exponent \cite{Collet,Hilborn}
\begin{equation}
\lambda = \lim _{n\rightarrow\infty} \frac{1}{n} \sum_{k=0}^{n-1}
\ln \Big|\frac {dx_{k+1}}{dx_k}\Big|\,,
\end{equation}
where $x_k = K_{ij}/\overline{K}$ at step $k$ of the renormalization-group trajectory.  We obtain the Lyapunov exponents by numerically calculating the logarithmic sum in Eq.(5), eliminating the first 100 renormalization-group steps as crossover from initial conditions to asymptotic behavior and then using the next 1500 steps.  This calculation, for all spins $s$, yields $\lambda = 1.93$, numerically correct to the three shown significant figures.

In addition to strong chaos, the renormalization-group trajectories show asymptotic strong coupling behavior,
\begin{equation}
\overline{K'} = b^{y_R}\, \overline{K}\,,
\end{equation}
where $y_R >0$ is the runaway exponent \cite{Demirtas}.  Again using 1500 renormalization-group steps after discarding 100 steps, we find $y_R = 0.24$ for all spins $s$, numerically correct to the two shown significant figures. We thus find that $y_R$ and $\lambda$ are universal with respect to spin $s$.

Note that here is a "weak" strong coupling behavior, as the stronger runaway exponent of the ferromagnetic and antiferromagnetic phases is $y_R = d-1 = 2$.  In fact, the runaway exponent $y_R$  occurs, under renormalization group, in all ordered phases and is $y_R= d-1$ in all conventionally ordered phases.  However, in the spin-glass phase, since deep into this phase where the renormalization-goup flows lead (at the sink of this phase), we do not have the rescaling of compactly ordered renormalization-group cells \cite{McKayChaos}, $y_R<d-1$, as also seen in Refs.\cite{Demirtas,Atalay}, where the runaway exponents $y_R$ have been studied in detail.
\section{Conclusion}

We have calculated the global spin-$s$ sequence of phase diagrams for all spins $s=1/2,1,3/2,2,5/2,3,...,s\rightarrow\infty$ for the Ising spin-glass system in spatial dimension $d=3$.  The phase diagrams, all with a finite-temperature spin-glass phase, for increasing spin $s$ saturate to the limit value of $s\rightarrow\infty$.  For all spins $s$, the spin-glass phase has renormalization-group trajectories that are chaotic, with calculated Lyapunov exponent $\lambda = 1.93$ and runaway exponent $y_R=0.24$, thus simultaneously showing strong chaotic and "weak" strong-coupling behaviors.

\begin{acknowledgments}
Support by the Kadir Has University Doctoral Studies Scholarship Fund and by the Academy of Sciences of Turkey (T\"UBA) is gratefully acknowledged.
\end{acknowledgments}


\begin{references}

\bibitem{spinS1} C. Domb and M. F. Sykes, Effect of Change of Spin on the Critical Properties of the Ising and Heisenberg Models, Phys. Rev. {\bf 128}, 168 (1962).
\bibitem{spinS2} Phase transitions and critical phenomena: Series expansions for lattice models, Vol. 3, C. Domb and M. S. Green, eds. (London, New York, Academic Press, 1974).
\bibitem{spinS3} J. P. Vandyke and W. J. Camp, High-temperature series for susceptibility of spin-s Ising model, Phys. Rev. B {\bf 9}, 3121  (1974).
\bibitem{spinS4} W. J. Camp and J. P. Vandyke, High-temperature series for susceptibility of spin-s Ising model: Analysis of  confluent singularities, Phys. Rev. B {\bf 11}, 2579  (1975).
\bibitem{spinS5} A. N. Berker, Critical Interactions for the Triangular Spin-s Ising Model by a Spin-Restructuring Transformation, Phys. Rev. B {\bf 12}, 2752 (1975).
\bibitem{spinS6} W. J. Camp, D. M. Saul, J. P. Vandyke, and M. Wortis, Series analysis of corrections to scaling for spin-pair correlations of spin-s Ising model: Confluent singularities, universalityi and hyperscaling, Phys. Rev. B {\bf 14}14, 3990 (1976).
\bibitem{spinS7} B. Nienhuis, A.N. Berker, E.K. Riedel, and M. Schick, First- and Second-Order Phase Transitions in Potts Models: Renormalization-Group Solution, Phys. Rev. Lett. {\bf 34}, 737 (1979).
\bibitem{spinS8} I. Jensen, A. J. Guttmann, and I. G. Enting, Low-temperature series expansions for the square lattice Ising model with spin S>1, J. Phys. A {\bf 29}, 3805 (1996).
\bibitem{spinS9} S. Sardar and K. G. Chakraborty,  Analysis of coherent anomalies, scaling exponent and confluent singularities for spin-S Ising model on cubic nets, Physica A {\bf 238}, 317 (1997)
\bibitem{spinS10} T. Yokota, On the ground state of Ising models with arbitrary spin quantum number, Phys. Lett. A {\bf 298}, 236 (2002).
\bibitem{spinS11} P. Butera and M. Comi, Critical universality and hyperscaling revisited for Ising models of general spin using extended high-temperature series, Phys. Rev. B {\bf 65}, 144431 (2002)
\bibitem{spinS12} P. Butera and M. Comi, An on-line library of extended high-temperature expansions of basic observables for the spin-S Ising models on two- and three-dimensional lattices, J. Stat. Phys. {\bf 109}, 311 (2002).
\bibitem{spinS13} P. Butera, M. Comi, and A. J. Guttmann, Critical parameters and universal amplitude ratios of two-dimensional spin-S Ising models using high- and low-temperature expansions, Phys. Rev. B {\bf 67}, 054402 (2003).
\bibitem{spinS14} P. Butera and M. Comi, Updated tests of scaling and universality for spin-spin correlations in the two- and three-dimensional spin-S Ising models using high-temperature expansions, Phys. Rev. B {\bf 69}, 174416 (2004).
\bibitem{spinS15} O. Rojas, S. M.de Souza, and W. A. Moura-Melo, On the series expansion of the general spin-s Ising chain, Physica A {\bf 373}, 324 (2007).

\bibitem{Aizenman} M. Aizenman and J. Wehr, Rounding of First-Order Phase Transitions in Systems with Quenched Disorder, Phys. Rev. Lett. {\bf 62}, 2503 (1989).
\bibitem{AizenmanE} M. Aizenman and J. Wehr, Phys. Rev. Lett. {\bf 64}, 1311(E) (1990).
\bibitem{HuiBerker} K. Hui and A. N. Berker, Random-Field Mechanism in Random-Bond Multicritical Systems, Phys. Rev. Lett. {\bf 62}, 2507 (1989).
\bibitem{erratum} K. Hui and A. N. Berker, erratum, Phys. Rev. Lett. {\bf 63}, 2433 (1989).
\bibitem{EdwardsAnderson} S.F. Edwards and P. W. Anderson, Theory of Spin Glasses, J. Phys. F {\bf 5}, 965 (1975).
\bibitem{Toulouse} G. Toulouse, Theory of Frustration Effect in Spin Glasses 1., Commun. Phys. {\bf2}, 115 (1977).

\bibitem{McKayChaos} S. R. McKay, A. N. Berker, and S. Kirkpatrick, Spin-Glass Behavior in Frustrated Ising Models with Chaotic Renormalization-Group Trajectories, Phys. Rev. Lett. {\bf 48}, 767 (1982).
\bibitem{McKayChaos2} S. R. McKay, A. N. Berker, and S. Kirkpatrick, Amorphously Packed, Frustrated Hierarchical Models: Chaotic Rescaling and Spin-Glass Behavior, J. Appl. Phys. {\bf 53}, 7974 (1982).
\bibitem{BerkerMcKay} A. N. Berker and S. R. McKay, Hierarchical Models and Chaotic Spin Glasses, J. Stat. Phys. {\bf 36}, 787 (1984).
\bibitem{Hartford} E. J. Hartford and S. R. McKay, Ising Spin-Glass Critical and Multicritical Fixed Distributions from a Renormalization-Group Calculation with Quenched Randomness, J. Appl. Phys. {\bf 70}, 6068 (1991).

\bibitem{ZZhu}Z. Zhu, A. J. Ochoa, S. Schnabel, F. Hamze, and H. G. Katzgraber, Best-Case Performance of Quantum Annealers on Native Spin-Glass Benchmarks: How Chaos Can Affect Success Probabilities, Phys. Rev. A {\bf 93}, 012317 (2016).
\bibitem{Katzgraber3}W. Wang, J. Machta, and H. G. Katzgraber, Bond Chaos in Spin Glasses Revealed through Thermal Boundary Conditions, Phys. Rev. B {\bf 93}, 224414 (2016).
\bibitem{Fernandez} L. A. Fernandez, E. Marinari, V. Martin-Mayor, G. Parisi, and D. Yllanes, Temperature Chaos is a Non-Local Effect, J. Stat. Mech. - Theory and Experiment, 123301 (2016).
\bibitem{Fernandez2} A. Billoire, L. A. Fernandez, A. Maiorano, E. Marinari, V. Martin-Mayor, J. Moreno-Gordo, G. Parisi, F. Ricci-Tersenghi, J.J. Ruiz-Lorenzo, Dynamic Variational Study of Chaos: Spin Glasses in Three Dimensions, J. Stat. Mech. - Theory and Experiment, 033302 (2018).
\bibitem{Wang2} W. Wang, M. Wallin, and J. Lidmar, Chaotic Temperature and Bond Dependence of Four-Dimensional Gaussian Spin Glasses with Partial Thermal Boundary Conditions, Phys. Rev. E {\bf98}, 062122 (2018).
\bibitem{Eldan} R. Eldan, The Sherrington-Kirkpatrick Spin Glass Exhibits Chaos, J. Stat. Phys. {\bf 181}, 1266 (2020).
\bibitem{Parisi3} M. Baity-Jesi, E. Calore, A. Cruz, L. A. Fernandez, J. M. Gil-Narvion, I. G.-A. Pemartin, A. Gordillo-Guerrero, D. I\~{n}iguez, A. Maiorano, E. Marinari, V. Martin-Mayor, J. Moreno-Gordo, A. Mu\~{n}oz-Sudupe, D. Navarro, I. Paga, G. Parisi, S. Perez-Gaviro, F. Ricci-Tersenghi, J. J. Ruiz-Lorenzo, S. F. Schifano, B. Seoane, A. Tarancon, R. Tripiccione, and D. Yllanes, Temperature Chaos Is Present in Off-Equilibrium Spin-Glass Dynamics, Comm. Phys. {\bf 4}, 74 (2021).

\bibitem{Aral} N. Aral and A. N. Berker, Chaotic Spin Correlations in Frustrated Ising Hierarchical Lattices, Phys. Rev. B {\bf 79}, 014434 (2009).

\bibitem{Mattis} D. C. Mattis, Solvable Spin Systems with Random Interactions, Phys. Lett. A {\bf 56}, 421 (1976).
\bibitem{Ilker2} E. Ilker and A. N. Berker, Overfrustrated and Underfrustrated Spin Glasses in d=3 and 2: Evolution of Phase Diagrams and Chaos including Spin-Glass Order in d=2, Phys. Rev. E {\bf 89}, 042139 (2014).

\bibitem{Parisi} S. Franz, G. Parisi, and M.A. Virasoro, Interfaces and Lower Critical Dimension in a Spin-Glass Model, J. Physique I {\bf 4}, 1657 (1994).
\bibitem{Amoruso} C. Amoruso, E. Marinari, O. C. Martin, and A. Pagnani, Scalings of Domain Wall Energies in Two Dimensional Ising Spin Glasses, Phys. Rev. Lett. {\bf 91}, 087201 (2003).
\bibitem{Bouchaud} J.-P. Bouchaud, F. Krzakala, and O. C. Martin, Energy Exponents and Corrections to Scaling in Ising Spin Glasses, Phys. Rev. B {\bf 68}, 224404 (2003).
\bibitem{Boettcher} S. Boettcher, Stiffness of the Edwards-Anderson Model in All Dimensions, Phys. Rev. Lett. {\bf95}, 197205 (2005).
\bibitem{Demirtas} M. Demirtaş, A. Tuncer, and A. N. Berker, Lower-Critical Spin-Glass Dimension from 23 Sequenced Hierarchical Models, Phys. Rev. E 92, 022136 (2015).
\bibitem{Parisi2} A. Maiorano and G. Parisi, Support for the Value 5/2 for the Spin Glass Lower Critical Dimension at Zero Magnetic Field, Proc. Natl. Acad. Sci. USA {\bf 115}, 5129 (2018).
\bibitem{Atalay} B. Atalay and A. N. Berker, A Lower Lower-Critical Spin-Glass Dimension from Quenched Mixed-Spatial-Dimensional Spin Glasses, Phys. Rev. E 98, 042125 (2018).

\bibitem{Grinstein} G. Grinstein, A. N. Berker, J. Chalupa, and M. Wortis, Exact Renormalization Group with Griffiths Singularities and Spin-Glass Behavior:  The Random Ising Chain, Phys. Rev. Lett. {\bf 36}, 1508 (1976).
\bibitem{Binder} I. Morgenstern and K. Binder, Evidence against Spin-Glass Order in the 2-Dimensional Random-Bond Ising Model, Phys. Rev. Lett. 43, 1615 (1979).
\bibitem{Ilker1} E. Ilker and A. N. Berker, High q-State Clock Spin Glasses in Three Dimensions and the Lyapunov Exponents of Chaotic Phases and Chaotic Phase Boundaries, Phys. Rev. E {\bf 87}, 032124 (2013).
\bibitem{Ilker3} E. Ilker and A. N. Berker, Odd q-State Clock Spin-Glass Models in Three Dimensions, Asymmetric Phase Diagrams, and Multiple Algebraically Ordered Phases, Phys. Rev. E {\bf 90}, 062112 (2014).

\bibitem{Surface3} M. Kardar and A. N. Berker, Commensurate-Incommensurate Phase Diagrams for Overlayers from a Helical Potts Model, Phys. Rev. Lett. {\bf 48}, 1552 (1982).

\bibitem{Caglar1} T. \c{C}a\u{g}lar and A. N. Berker, Chiral Potts Spin Glass in d = 2 and 3 Dimensions, Phys. Rev. E {\bf 94}, 032121 (2016).
\bibitem{Caglar2} T. \c{C}a\u{g}lar and A. N. Berker, Devil’s Staircase Continuum in the Chiral Clock Spin Glass with Competing
Ferromagnetic-Antiferromagnetic and Left-Right Chiral Interactions, Phys. Rev. E {\bf 95}, 042125 (2017).
\bibitem{Caglar3} T. \c{C}a\u{g}lar and A. N. Berker, Phase Transitions Between Different Spin-Glass Phases and Between Different Chaoses
in Quenched Random Chiral Systems, Phys. Rev. E {\bf 96}, 032103 (2017).
\bibitem{Kaplan} C. N. Kaplan and A. N. Berker, Quantum-Mechanically Induced Asymmetry in the Phase Diagrams of Spin-Glass Systems, Phys. Rev. Lett. 100, 027204, 1-4 (2008).

\bibitem{AndelmanBerker} D. Andelman and A. N. Berker, Scale-Invariant Quenched Disorder and its Stability Criterion at Random Critical Points, Phys. Rev. B {\bf 29}, 2630 (1984).

\bibitem{Migdal} A. A. Migdal, Phase Transitions in Gauge and Spin Lattice Systems, Zh. Eksp. Teor. Fiz. {\bf69}, 1457 (1975) [Sov. Phys. JETP {\bf42}, 743 (1976)].
\bibitem{Kadanoff} L. P. Kadanoff, Notes on Migdal's Recursion Formulas, Ann. Phys. (N.Y.) {\bf100}, 359 (1976).
\bibitem{Biroli} M. C. Angelini and G. Biroli, Real Space Migdal-Kadanoff Renormalisation of Glassy Systems: Recent Results and a Critical
Assessment, J. Stat. Phys. {\bf167}, 476 (2017).
\bibitem{BerkerOstlund} A. N. Berker and S. Ostlund, Renormalisation-Group Calculations of Finite Systems: Order Parameter and Specific Heat for Epitaxial Ordering, J. Phys. C {\bf 12}, 4961 (1979).
\bibitem{Kaufman1} R. B. Griffiths and M. Kaufman, Spin Systems on Hierarchical Lattices: Introduction and Thermodynamic Limit, Phys. Rev. B {\bf 26}, 5022R (1982).
\bibitem{Kaufman2} M. Kaufman and R. B. Griffiths, Spin Systems on Hierarchical Lattices: 2. Some Examples of Soluble Models, Phys. Rev. B {\bf 30}, 244 (1984).
\bibitem{ArtunBerker} E. C. Artun and A. N. Berker, Complete Density Calculations of q-State Potts and Clock Models: Reentrance of Interface Densities under Symmetry Breaking, Phys. Rev. E {\bf 102}, 062135 (2020).

\bibitem{Myshlyavtsev} A. V. Myshlyavtsev, M. D. Myshlyavtseva, and S. S. Akimenko, Classical Lattice Models with Single-Node Interactions on Hierarchical Lattices: The Two-Layer Ising Model, Physica A {\bf558}, 124919 (2020).
\bibitem{Derevyagin} M. Derevyagin, G. V. Dunne, G. Mograby, and A. Teplyaev, Perfect Quantum State Transfer on Diamond Fractal Graphs, Quantum Information Processing, {\bf19}, 328 (2020).
\bibitem{Shrock} S.-C. Chang, R. K. W. Roeder, and R. Shrock, q-Plane Zeros of the Potts Partition Function on Diamond Hierarchical Graphs, J. Math. Phys. {\bf61}, 073301 (2020).
\bibitem{Monthus} C. Monthus, Real-Space Renormalization for Disordered Systems at the Level of Large Deviations, J. Stat. Mech. - Theory and Experiment, 013301 (2020).
\bibitem{Sariyer} O. S. Sar{\i}yer, Two-Dimensional Quantum-Spin-1/2 XXZ Magnet in Zero Magnetic Field: Global Thermodynamics from Renormalisation Group Theory, Philos. Mag. {\bf 99}, 1787 (2019).
\bibitem{Ruiz} P. A. Ruiz, Explicit Formulas for Heat Kernels on Diamond Fractals, Comm. Math. Phys. {\bf 364}, 1305 (2018).
\bibitem{Rocha-Neto} M. J. G. Rocha-Neto,  G. Camelo-Neto, E. Nogueira, Jr., and S. Coutinho, The Blume–Capel Model on Hierarchical Lattices: Exact Local Properties, Physica A {\bf494}, 559 (2018).
\bibitem{Ma} F. Ma, J. Su, Y. X. Hao, B. Yao, and G. G. Yan, A Class of Vertex–Edge-Growth Small-World Network Models Having Scale-Free, Self-Similar and Hierarchical Characters Physica A {\bf492}, 1194 (2018).
\bibitem{Boettcher5} S. Boettcher and S. Li, Analysis of Coined Quantum Walks with Renormalization, Phys. Rev. A {\bf 97}, 012309 (2018).

\bibitem{Collet} P. Collet and J.-P. Eckmann, \textit{Iterated Maps on the Interval as Dynamical Systems} (Birkh\"{a}user, Boston, 1980).
\bibitem{Hilborn} R. C. Hilborn, \textit{Chaos and Nonlinear Dynamics}, 2nd ed. (Oxford University Press, New York, 2003).

\end{references}
\end{document}